\documentclass[10pt,a4paper,final]{article}
\usepackage[latin1]{inputenc}
\usepackage{amsmath}
\usepackage{amsfonts}
\usepackage{amssymb}
\usepackage{graphics}

\title{Generalization of Conway's "Game of Life" to a continuous domain - SmoothLife}

\author{
        Stephan Rafler \\
        N\"urnberg, Germany\\
        frlndmr@web.de
}

\date{\today}

\begin{document}

\maketitle

\begin{abstract}

We present what we argue is the generic generalization of Conway's "Game of Life" to a continuous domain. We describe the theoretical model and the explicit implementation on a computer.

\end{abstract}

\section{Introduction}

There have been many generalizations of Conway's "Game of Life" (GoL) since its invention in 1970 \cite{cgl}. Almost all attributes of the GoL can be altered: the number of states, the grid, the number of neighbors, the rules. One feature of the original GoL is the glider, a stable structure that moves diagonally on the underlying square grid. There are also "spaceships", similar structures that move horizontally or vertically.

Attempts to construct gliders (as we will call all such structures in the following), that move neither diagonally nor straight, have led to huge man-made constructions in the original GoL. An other possibility to achieve this has been investigated by Evans \cite{ltl}, namely the enlargement of the neighborhood. It has been called "Larger than Life" (LtL). Instead of 8 neighbors the neighborhood is now best described by a radius $r$, and a cell having $ (2r+1)^2-1 $ neighbors. The rules can be arbitrarily complex, but for the start it is sensible to consider only such rules that can be described by two intervals. They are called "birth" and "death" intervals and are determined by two values each. These values can be given explicitly as the number of neighbors or by a filling, a real number between 0 and 1. In the first case, the radius has to be given, too, in the last case, this can be omitted. 

The natural extension of Evans' model is to let the radius of the neighborhood tend to infinity and call this the continuum limit. The cell itself becomes an infinitesimal point in this case. This has been done by Pivato \cite{rel} and investigated mathematically. He has called this model "RealLife" and has given a set of "still lives", structures that do not evolve with time.

\section{SmoothLife}\label{smth}

We take a slightly different approach and let the cell not be infinitesimal but of a finite size. Let the form of the cell be a circle (disk) in the following, although it could be any other closed set. Then, the "dead or alive" state of the cell is not determined by the function value at a point $\vec x$, but by the filling of the circle around that point. Similarly, the filling of the neighborhood is considered. Let the neighborhood be ring shaped, then with $f(\vec x,t)$ our state function at time $t$ we can determine the filling of the cell or "inner filling" $m$ by the integral

\begin{equation}
m = \frac{1}{M} \int\limits_{|\vec u|<r_i} \! d \vec u \, f(\vec x + \vec u,t)
\end{equation}

and the neighborhood or "outer filling" $n$ by the integral

\begin{equation}
n = \frac{1}{N} \int\limits_{r_i<|\vec u|<r_a} \! d \vec u \, f(\vec x + \vec u,t)
\end{equation}

where $N$ and $M$ are normalization factors such that the filling is between 0 and 1. Because the function values of $f$ lie also between 0 and 1 the factors simply consist of the respective areas of disk and ring. The radius of the disk or "inner radius" is given by $r_i$ which is also the inner radius of the ring. The outer radius of the ring is given by $r_a$.

In the original GoL the state of a cell for the next time-step is determined by two numbers: the live-state of the cell itself, which is 0 or 1, and the number of live neighbors, which can be between 0 and 8. One could model all general rules possible by a $2 \times 9$ matrix containing the new states for the respective combinations. It could be called the transition matrix.

Now in our case this translates to the new state of the point $\vec x$ being determined by the two numbers $m$ and $n$. The new state is given by a function $s(n,m)$. Let us call it the transition function. It is defined on the interval $[0,1] \times [0,1]$ and has values in the range $[0,1]$. To resemble the corresponding situation in GoL, typically $r_a = 3 r_i$ is chosen (the diameter of the neighborhood is 3 cells wide).

\section{Computer implementation}\label{comp}

As simple as the theoretical model is, it is not immediately obvious, how to implement it on a computer, as a computer cannot handle infinitesimal values, continuous domains, etc. But it can handle real numbers in the form of floating point math, and as it turns out, this is sufficient. We also can model the continuous domain by a square grid, the ideal data structure for computation. So we will be able to implement our function $f(\vec x,t)$ as a $\mathsf{float}$ array.

When implementing the circularly shaped integrals we run into a problem. Pixelated circles typically have jagged rims. So either we let the radius of the circle be so huge, that the pixelation due to our underlying square grid is negligible. Then the computation time will be enormous. Or we use another solution used in many similar situations: anti-aliasing. Consider for example the integration of the inner region. For the cell $\vec x$ function values are taken at locations $\vec x + \vec u$. Let us define $l = |\vec u|$. With an anti-aliasing zone around the rim of width $b$ we take the function value as it is, when $l<r_i-b/2$. In the case when $l>r_i+b/2$ we take 0. In between we multiply the function value by $(r_i+b/2-l)/b$. Similarly for the inner rim of the ring and the outer rim. In this way the information on how far the nearest grid point is away from the true circle, is retained. Typically, $b=1$ is chosen.

We also have to construct the transition function $s(n,m)$ explicitly. Luckily we can restrict ourselves like LtL, for the beginning, to four parameters: the boundaries of the birth and death intervals. To make things smooth and to stay in the spirit of the above described anti-aliasing we use smooth step functions instead of hard steps. We call them sigmoid functions to emphasize this smoothness. For example we could define

\begin{equation}
\sigma_1(x,a) = \frac{1}{1 + \exp(-(x-a)4/\alpha)}
\end{equation}

\begin{equation}
\sigma_2(x,a,b) = \sigma_1(x,a) \ (1 - \sigma_1(x,b))
\end{equation}

\begin{equation}
\sigma_m(x,y,m) = x (1 - \sigma_1(m, 0.5)) + y \, \sigma_1(m, 0.5)
\end{equation}

then we can define the transition function as

\begin{equation}
s(n,m) = \sigma_2 (n, \sigma_m(b_1, d_1, m), \sigma_m(b_2, d_2, m))
\end{equation}

where birth and death intervals are given by $[b_1,b_2]$ and $[d_1,d_2]$ respectively. The width of the step is given by $\alpha$. As we have two different types of steps we have an $\alpha_n$ and an $\alpha_m$. Note that neither the anti-aliasing nor the smoothness of the transition function are necessary for the computer simulation to work. They just make things smoother and allow one to choose smaller radii for neighborhood and inner region and so achieve faster computation times for the time-steps.

\section{Smooth time-stepping}\label{stime}

So far we have made everything smooth and continuous but one thing: the time-steps are still discrete. At time $t$ the function $s(n,m)$ is calculated for every $\vec x$ and this gives the new value $f(\vec x,t+1)$ at time $t+1$. If we think of the application of $s(n,m)$ as a nonlinear operator $S$ we can write

\begin{equation}
f(\vec x,t+1) = S[s(n,m)] \, f(\vec x,t)
\end{equation}

To give us the ability to obtain arbitrarily small time steps, we introduce an infinitesimal time $dt$ and reinterpret the transition function as a rate of change of the function $f(\vec x,t)$ instead of the new function value. Then we can write

\begin{equation}
f(\vec x,t+dt) = f(\vec x,t) + dt \, S[s(n,m)] \, f(\vec x,t)
\end{equation}

where we have defined a new $s(n,m)$, that has values in the range $[-1,1]$ instead of $[0,1]$. If the transition function in the discrete time-stepping scheme was $s_{d}(n,m)$ then the smooth one is $s(n,m) = 2 s_{d}(n,m) - 1$. The formula above is also the most trivial integration scheme for the integro-differential equation

\begin{equation}
\partial_t f(\vec x,t) = S[s(n,m)] \, f(\vec x,t)
\end{equation}

This equation however leads to a different form of life. The same generic gliders can not be found at the same birth/death values as in the version with discrete time-stepping, but it also leads to gliders, oscillating and stable structures.

\section{Conclusions}\label{conc}

We have described a model to generalize Conway's "Game of Life" to a continuous range of values and a continuous domain. The $2 \times 9$ transition matrix of the GoL has been generalized to the $s(n,m)$ transition function. The 8 pixel neighborhood and 1 pixel cell of GoL have been generalized to a ring shaped neighborhood and a disk shaped cell. The rule set has been generalized to four real numbers: the boundaries of the birth and death intervals. The last remaining discrete attribute was the time-stepping. We proposed a method for continuous time-stepping which reinterprets the transition function as the velocity of change.

The technique with two radii has been used in other contexts \cite{yan}, but no gliders were described. There has also been a computer implementation of a continuous version of GoL, but without the inner radius technique, and no gliders were found at that time \cite{len}.

The goal of finding a glider that can move in arbitrary directions has been achieved. Of the original GoL it resembles both the glider and the spaceship at the same time. It also resembles similar structures found in LtL. So we think we have found the generic, generalized glider, and call it the "smooth glider".

\begin{figure}[h]
\includegraphics{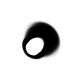} 
\caption{the smooth glider for $r_a=21,\ b_1=0.278,\ b_2=0.365,\ d_1=0.267,\ d_2=0.445,\  \alpha_n=0.028,\  \alpha_m=0.147$ as it moves to the upper right}
\end{figure}

\end{document}